\documentclass[monochrome]{elsarticle}

\usepackage[monochrome]{xcolor}
\usepackage{amssymb}
\usepackage{amsmath}
\usepackage{ifthen}
\usepackage{epsfig}
\usepackage{url}
\usepackage{float}
\usepackage{nomencl}
\usepackage[nolist]{acronym}
\usepackage{multicol}
\usepackage{bm}
\usepackage[numbers]{natbib}
\usepackage{txfonts}
\usepackage{graphicx}
\newcommand{\EQ}{\begin{equation}}
\newcommand{\EN}{\end{equation}}
\newcommand{\EQA}{\begin{eqnarray}}
\newcommand{\ENA}{\end{eqnarray}}

\newcommand{\mean}[1]{\overline #1}

\def\Rey{\mbox{\rm Re}}

\def\St{\mbox{\rm St}}
\def\Sh{\mbox{\rm Sh}}

\def\Da{\mbox{\rm Da}}

\newcommand{\uu}{\mbox{\boldmath $u$} {}}

\newcommand{\vv}{\mbox{\boldmath $v$} {}}

\newcommand{\ff}{\mbox{\boldmath $f$} {}}

\newcommand{\FF}{\mbox{\boldmath $F$} {}}

\newcommand{\SSS}{\mbox{\boldmath $S$} {}}

\newcommand{\nab}{\mbox{\boldmath $\nabla$} {}}

\def\Rey{{\rm Re}}

\def\nab{{\bm{\nabla}}}
\def\ff{{\bm f}}

\newcommand{\bez}{\begin{eqnarray*}}
\newcommand{\eez}{\end{eqnarray*}}
\newcommand{\be}{\begin{equation}}
\newcommand{\beq}{\begin{eqnarray}}
\newcommand{\eeq}{\end{eqnarray}}
\newcommand{\bc}{\begin{center}}
\newcommand{\ec}{\end{center}}

\usepackage[switch,modulo]{lineno}
\usepackage{titlesec}
\usepackage[top=2cm,bottom=4cm,left=3.5cm,right=3.5cm]{geometry}

\renewcommand*\nompreamble{\begin{multicols}{2}}
\renewcommand*\nompostamble{\end{multicols}}
\begin{document}
\small
\baselineskip 10pt
\begin{acronym}
 \acro{IGCC}{Integrated Gasification Combined Cycle}
 \acro{CCS}{Carbon Capture and Storage}
 \acro{RSM}{Reynolds stress modeling}
 \acro{CFD}{Computational Fluid Dynamics}
 \acro{DNS}{Direct Numerical Simulation}
 \acro{RANS}{Reynolds-averaged Navier Stokes}
 \acro{LES}{Large Eddy Simulation}
 \acro{CAD}{Computer Aided Design}
 \acro{EBU}{Eddy Break-Up}
\end{acronym}

\begin{frontmatter}

\title{The effect of turbulent clustering on particle reactivity}
\author[NTNU]{Jonas Kr\"uger\corref{correspondingauthor}}
\cortext[correspondingauthor]{Corresponding author}
\ead{jonas.kruger@ntnu.no}
\author[NTNU,Sintef]{Nils E. L. Haugen}
\author[Nordita]{Dhrubaditya Mitra}
\author[NTNU]{Terese L{\o}v{\aa}s}

\address[NTNU]{EPT, NTNU, 
    Kolbj{\o}rn Hejes vei 1B, N-7491 Trondheim, Norway}
\address[Sintef]{SINTEF Energy Research, N-7465 Trondheim, Norway}
\address[Nordita]{NORDITA, KTH Royal Institute of Technology and Stockholm 
University, Roslagstullsbacken 23, SE-10691 Stockholm, Sweden}

\begin{abstract}
The effect of turbulence on the heterogeneous (solid-fluid) 
reactions of solid
particles is studied numerically with Direct Numerical Simulations (DNS). 
A simplified reaction system is used, 
where the solid-fluid reaction is represented by a single isothermal 
reaction step. It is found that, due to the
clustering of particles by the isotropic turbulence, the
overall reaction rate is entirely controlled by the turbulence for 
large Damk\"ohler numbers. The particle clustering significantly slows down the
reaction rate for increasing Damk\"ohler numbers which reaches an asymptotic 
limit that can be analytically derived. 
This implies that the effect of turbulence on 
heterogeneously reacting particles should be included in models that are
used in CFD simulations of e.g. char burnout in combustors or gasifiers. 
Such a model, based on the chemical and turbulent time scales, is here
proposed for the heterogeneous reaction rate in the presence of 
turbulence.
\end{abstract}

\begin{keyword}
Turbulent reacting multiphase flow, char oxidation, clustering
\end{keyword}

\end{frontmatter}

\begin{multicols}{2}
\section{Introduction}
Particles that are exchanging mass with a surrounding turbulent flow fluid are 
found in a wide range of situations, both in nature and in industrial 
applications. Examples of these are pulverized coal combustion in large power 
plants and fluidized beds in the process industry. A general feature among all 
these cases is the multi-scale nature of the problem, where the smallest scale 
is typically the size of the particle, or even the internal structure of the 
particle, while the largest scale is the much larger size of the entire 
combustion chamber or reactor. In the intermediate range between these two 
extremes, one find the scales of the turbulence, which goes from the Kolmogorov 
scale to the energy carrying scale (the integral scale). Another feature common 
among these cases is that the particles exchange mass with the surrounding 
fluid 
through chemical reactions on the surface of the particles, as e.g. during 
oxidation or gasification of char. 

The effect of turbulence on different large scale properties of the flow, such 
as turbulent viscosity, diffusivity and conductivity has been known for a long 
time. A relatively large number of models have been developed in order to 
account for 
these effects, such as e.g. the k-$\varepsilon$ model \citep{Jones1972} and 
different 
versions of the Reynolds Stress Models (e.g. \citep{pope}). When homogeneous 
combustion is considered, relatively good models such as the Eddy 
Dissipation Model \citep{Magnussen}, different variants of Probability Density 
Function models (e.g. \citep{Dopazo_1994}) or models based on conditioned 
parameters such as Conditional Moment Closure models (e.g. 
\citep{Klimenko_Bilger}) 
are being used. 
However, for heterogeneous particle reactions, i.e. where the reactions take 
place at the solid surfaces of the particles, no such model for the effect of 
turbulence on the chemical reactions has yet been formulated.

A range of efforts to simulate heterogeneous conversion systems under turbulent 
conditions have been 
made. Among recent work one can mention that of Silaen and Wang 
\citep{Silaen2010} who simulated an existing gasifier with Reynolds-averaged 
Navier Stokes (RANS) using different turbulence models for the continuum phase 
and compared their 
results with measurements. The turbulence effect on particles was included 
using 
a stochastic tracking scheme for the particles position, hence turbulence was 
not taken into account for the heterogeneous reaction rate or transport of gas 
phase species to the particles. Vascellari et al. \citep{Vascellari2015} ran 2D 
RANS simulations with kinetics calibrated to experiments and a detailed 
description of the heterogeneous reactions inside the particle via an 
effectiveness factor and solving directly for the species partial pressure at 
the particles surface by assuming local equilibrium. They compare their 
simulations with measurements from an industrial-scale gasifier and achieve 
good 
agreement. Yet, to the knowledge of the authors, only very few studies of 
combustion or gasification, 
where account is made for the effect of turbulence on the heterogeneous char 
conversion, are published, among these are the papers of Luo et al. 
\cite{Luo2012}, Brosh \& Chakraborty 
\cite{
brosh_2014} and Brosh et al. \cite{brosh_2015}. Here, the Direct Numerical 
Simulations (DNS) 
approach is utilized, where all turbulence scales are explicitly resolved on 
the 
computational grid, and hence the effect of the turbulence is 
implicitly accounted for.

Despite all the effort that has been put into the development of models for 
turbulent homogeneous combustion or gasification,
no good model has been proposed for turbulent heterogeneous combustion or 
gasification. This means that when particles that experience solid reactions 
such as during char oxidation (i.e. gas phase species react with the solid part 
of the particle, not the volatile part) are embedded in a turbulent flow, the 
turbulence is typically never taken into account in the simulations. The goal 
of this paper is to 
highlight the effect of turbulence on solid particles using DNS, and to develop 
a simple 
model for the influence of turbulence on reactive particles. 

\section{Implementation}
The direct numerical simulations are performed with the Pencil Code 
\citep{pencilpage}, a finite
difference code for compressible reactive flows that is fully parallelizable and
shows good weak scaling behaviour. 
It uses a sixth-order finite difference
scheme for spatial discretization and a memory-efficient third-order 
Runge-Kutta scheme \citep{Williamson1980} for temporal discretization.
The particles are treated in a Lagrangian manner and a 
cloud-in-cell method is used both to interpolate the fluid phase 
variables at the particles position and for the back  
reaction from the particles to the fluid.

\subsection{Fluid phase equations}
In order to isolate the effect of turbulence on reactivity alone, we consider a 
simplified case with only one reactive species, which is treated as a scalar 
field advected by the carrier fluid. This reactant is passive for the fluid 
flow 
and is assumed to react only with the solid phase in a catalytic manner. 
As a result, the reactant is converted on the surface of the particle, but no 
mass and energy is exchanged with the particle. For 
simplicity, the reaction is further assumed to be neither endothermic nor 
proceeds at a constant rate $\lambda$, which only depends on the surface area 
of the solid phase.
The equation describing the 
conservation of mass for the isothermal flow is
\EQ
\frac{{\rm D}\rho}{{\rm D} t} = -\rho\nab \cdot \uu,
\label{eq:cons_mass}
\EN
with $\rho$ and $\uu$ being the fluid density and velocity, respectively, 
and the advective derivative is given by
\EQ
\frac{{\rm D}}{{\rm D} t} = \frac{\partial}{\partial t} +\uu\cdot\nabla.
\EN
The equation for the conservation of momentum is
\EQ
\label{eq:cons_mom}
\rho\frac{{\rm D} {\uu}}{{\rm D} t}= \nabla\cdot (2\mu \SSS)
+ \rho\ff + \FF,
\EN
where  viscous effects are accounted for by the traceless rate of strain 
tensor $\SSS$ and the dynamic viscosity $\mu$.
To obtain statistically stationary isotropic turbulence, we employ a forcing
function $\ff$ equivalent to that of Babkovskaia et al. 
\citep{Babkovskaia2011}. The force is acting 
on the wavevectors lying on a shell in Fourier space with a radius of $k_f$, 
accelerating the flow at low wavenumbers. 
The flow 
integral scale is given by ${L}_f=L_xk_1/k_f$ when $L_x$ is the size of 
the simulation box and $k_1=1$~m$^{-1}$ is the wave number of the box size.
Since some cases simulated here have significant mass loading,
influencing the turbulence field \citep{Squires1990, Gore1989}, the term 
$\FF$ 
represents the drag force the particle exerts on the fluid phase.

The conservation equation for the molar fraction $X$ of the reactant
reads:
\EQ
\frac{\partial X}{\partial t}+ \nabla \cdot (X\uu)  =
- D \nabla^2 X + \frac{\hat{R}}{\rho},
\label{eq:cons_species}
\EN
with $D$ being the diffusivity of the reactant and $\hat{R}$ the source 
term due to the conversion of the reactant 
at the particles surface. 
Thus the reactant can be thought of as oxygen 
reacting with the carbon of an long lasting char particle 
without any thermal or flow effects.

\subsection{Particle equations}
The particles are modeled using a Lagrangian approach. They are spherical and 
treated as point
particles as the particle size is significantly smaller than the grid size. As 
the density of the particle is magnitudes higher than the fluid phase, we 
assume that the only force acting on the particle is the Stokes drag. Gravity 
forces are neglected to achieve a system that is independent of direction. The 
velocity $\vv$ of the particle is evolved as
\EQ
\frac{d \vv}{dt}= \frac{1}{\tau_p}(\uu-\vv)=\frac{\FF}{m_p},
\EN
with the particle stopping time given as $\tau_p= B d_p^2/18 \nu(1+f_c)$ when 
$f_c=0.15\Rey^{0.687}$ is due to
the Schiller-Naumann correlation \citep{schiller1933}. 
Here $B$ stands for the density 
ratio between particle and fluid, $d_p$ is the particles diameter and $m_p$ the 
particles mass.

\begin{figure}[H]
\centering
\includegraphics[width=.4\textwidth]{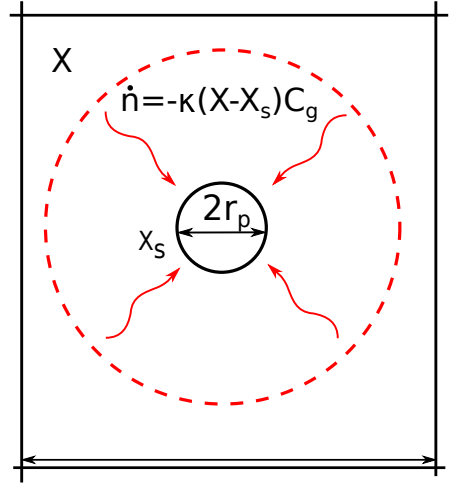}
\caption{Visualization of the flux to the particles surface}
\label{fig:diff}
\end{figure}

The reactant that is carried by the fluid phase is converted at the 
particle surface at a rate of $\hat{R}=A_p \dot{n} \bar{M} / V_{cell}$, where 
$A_p$ is 
the particles surface area, $\dot{n}$ is the reactant conversion rate per 
surface area, $\bar{M}$ the molar mass of the carrier fluid and 
$V_{cell}$ the volume of one grid cell. By letting the reactant molar 
fraction 
be denoted by $X_s$ at the particles surface and $X$ at a large 
distance from the surface, one can express the reactant conversion at the 
surface by $\dot{n} = - 
\lambda X_s C_g$, where $\lambda$ is the surface specific molar conversion 
rate. This is illustrated in Fig. \ref{fig:diff}. Assuming that the conversion 
at the surface is in equilibrium with the 
diffusive flux 
from the fluid phase to the particles 
surface $\dot{n}=-\kappa (X-X_s) C_g$, and solving for the surface 
mole fraction of the reactant, 
a modified reaction rate $\tilde{\lambda}$ for the reactant conversion
is defined as
\EQ
\dot{n}  = -\tilde{\lambda} X C_g
\EN
when
$\tilde{\lambda}=\lambda/(1+\lambda C_g/\kappa)$
is obtained following the ideas of Baum and Street \citep{Baum1971}. The 
adapted reaction rate $\tilde{\lambda}$ for the conversion rate will be 
important when formulating a model for the effect of turbulent clustering on 
the reactivity. The diffusion coefficient is here given by
$\kappa = D \Sh /2 r_p$ where the Sherwood coefficient $\Sh$ is set to two 
for all particle sizes representing quiescent flow around the particles for 
simplicity as the focus in this work is on macroscopic effects. 

\subsection{The limits of the decay rate}
In order to study the effect of turbulent clustering on the reactant conversion
it is useful to identify the governing chemical and turbulent time
and length scales.

The integral flow time scale $\tau_{L} ={L}_f / \uu_{rms}$ is based on 
the root mean square  of the velocity $u_{rms}$ and the scale of the forcing 
${L}_f$. 
It is often claimed that the 
highest value of the preferential concentration for particles is 
found for particles with a Stokes number around unity based on the Kolmogorov 
scale $\eta$ (\citep{Wood2005}, \citep{Eaton1994}). 
Since for the current study it is the large scale clusters that are interesting,
the Stokes number is here, however, defined based on the 
integral time scale
$\tau_{L}$ such that $\St = \tau_p / \tau_{L}$. 
Assuming 
homogeneously distributed particles, the instantaneous 
value of the reactant molar fraction $X$ is given by:
\EQ 
X(t) =  X_{0} e^{-n_p \tilde{\lambda} A_p t} =  X_{0} e^{-\alpha_{th} t}.
\EN
If the initial molar fraction of the reactant $X_{0}$ is set to unity, 
and 
the particle number density $n_p$ is given, the maximum theoretical decay rate  
$\alpha_{th}= n_p \tilde{\lambda} A_p$ can thus be estimated. Its 
inverse $1/\alpha_{th}=\tau_{th}$ is the theoretical reactive time scale. 

By defining a Damk\"ohler number $\Da= \tau_{L}/\tau_{th}$, the 
evolution of the decay rate $\alpha$ with Damk\"ohler number can be studied.
For low particle number densities, and therefore small 
Damk\"ohler numbers, 
the macroscopic clustering of particles can be neglected, and 
$\Da =\tau_{L}/\tau_{th}=\tau_{L} \alpha_{th}$
can be formed to yield the {\it particle} dependent decay rate 
$\alpha_{th} = \alpha_{p} = \Da / \tau_{L}$, which implies that the 
decay rate increases linearly with the Damk\"ohler number. For higher particle 
number densities, the macroscopic clusters have high internal particle 
number densities, which are rapidly converting the reactant within the cluster. 
Now the decay 
rate is controlled by the transport of reactant to the surface of these 
macroscopic clusters, and it is reasonable to consider the particle clusters as 
single bodies, or super-particles, that the reactants are converted at. One 
can 
then formulate a {\it cluster} dependent decay rate as $\alpha_c= n_c 
\tilde{\lambda}_c \bar{A}_c$, which is based on the cluster number density 
$n_c$, the modified conversion rate $\tilde{\lambda_c}$ and cluster 
surface area $\bar{A}_c$, which are constants dependent on the macroscopic flow 
field. 
One can now estimate values for $n_c$, $\tilde{\lambda}_c$ and $\bar{A}_c$, 
to find $\alpha_c$.
The following proposed formulation of $\alpha (\Da)$ then satisfies the 
limits as derived at high and low values of Damk\"ohler numbers:
\EQ
\alpha (\Da) = \frac{\alpha_c \Da}{\alpha_c \tau_{L} + \Da}.
\label{eq:alpha_Da}
\EN
This formulation will be compared directly to results from direct numerical 
simulations in the following section. 

\section{Results and discussion}
The computational domain for the DNS is a cube with an edge length of 
$2\pi$~cm, 
discretized 
with 64, 128 or 256 cells which results in grid cell sizes of 981, 490 and 
245 $\mu$m, respectively with increasing particle numbers. 
All boundaries 
are periodic. The strength of the forcing is chosen such that a 
turbulent Reynolds number 
$\Rey=u_{rms} {L}_f / \nu$ of approximately 250 is obtained for all cases.
Furthermore, the size and density of the particles are chosen to give
particle Stokes 
numbers of  $\St=0.1$ and $\St=1.0$.

\begin{figure}[H]
\centering\includegraphics[width=.45\textwidth]{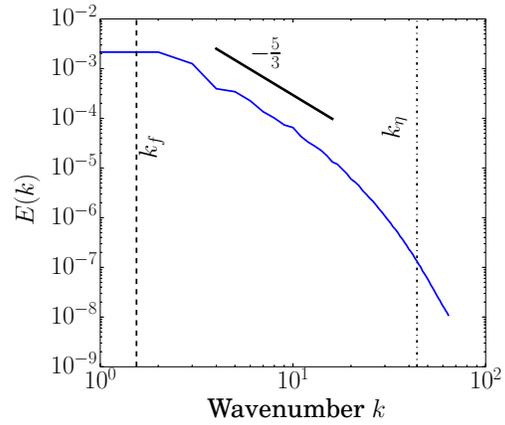}
\caption{
Power spectrum over the wavenumber for a case with a $\Rey\approx 250$ 
and a resolution of 128 cells. Shown are the forcing wavenumber $k_f$, the wave 
number of the Kolmogorov scale $k_\eta$ and the Kolmogorov scaling.}
\label{fig:spectrum}
\end{figure}

Figure \ref{fig:spectrum} shows the energy spectrum for a 
case with a Reynolds number of 250. The energy inserted into the domain at the 
forcing wavenumber $k_f$ (1.5 times the lowest wavenumber of the domain) is 
indeed transported to the higher wavenumbers via the inertial subrange,
which can be identified by its 
$-5/3$ slope, before it is dissipated around the Kolmogorov wavenumber $k_\eta$.

\begin{figure}[H]
\centering\includegraphics[width=.4\textwidth]{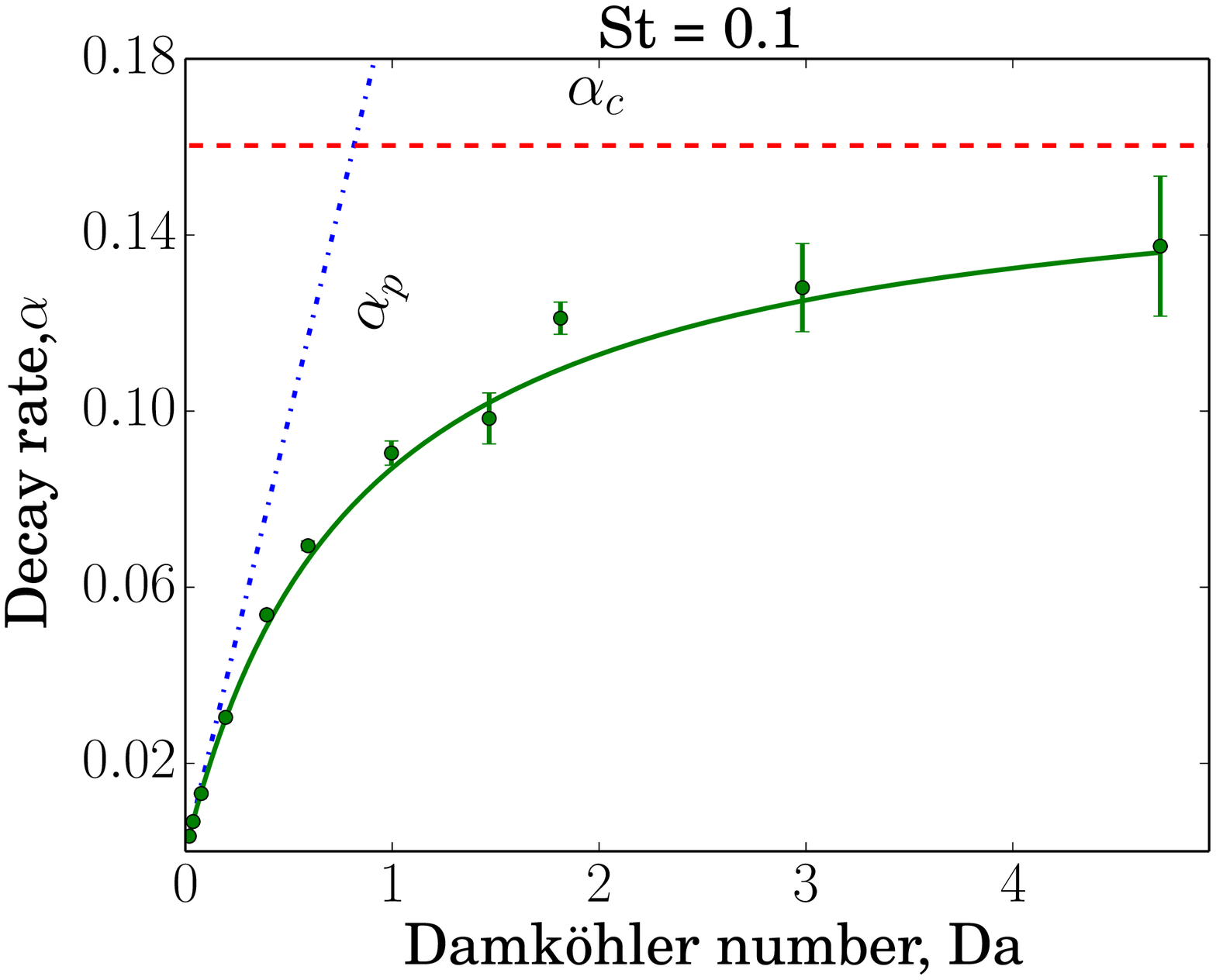}
\centering\includegraphics[width=.4\textwidth]{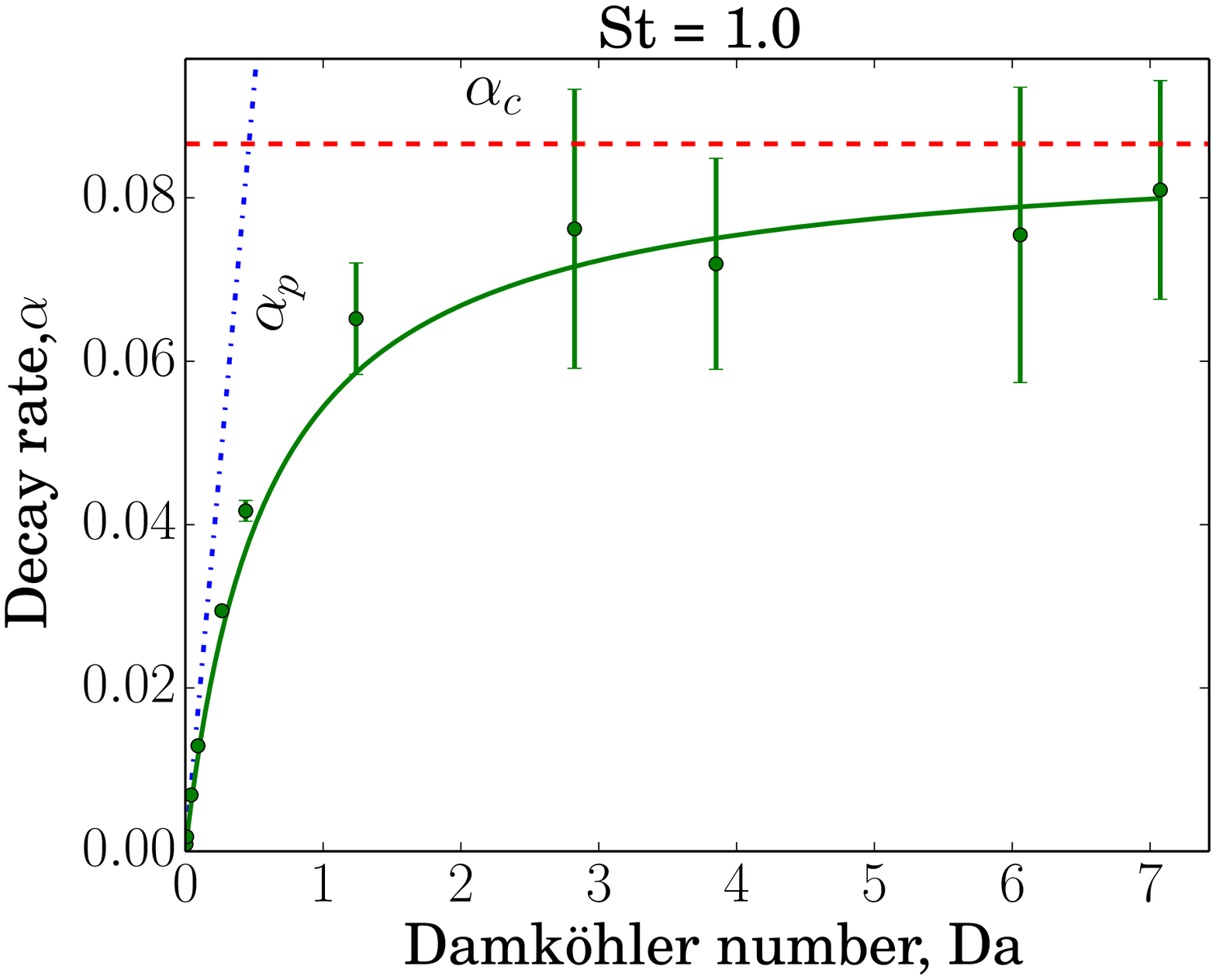}
\caption{Decay rate over Damk\"ohler number. The upper plots is for 
$\St_i\approx 0.1$, and the
lower for $\St_i\approx 1$. Filler circles represent the decay rates of the
numerical simulations, while the solid lines are the fits to the numerical 
results as given by Eq. \ref{eq:alpha_Da}. The dashed-dotted lines correspond 
to $\alpha_p$ while the dashed  lines represent $\alpha_c$.}
\label{fig:alpha_Da}
\end{figure}

The simulations are run until the turbulence is statistical stationary, then 
the 
molar fraction of the reactant is re-initialized to unity. 
The decay rate 
is obtained by fitting an exponential function to the reactant molar 
fraction from the start of data sampling to later times. 
The resulting decay rates as function of Damk\"ohler numbers are shown in Fig. 
\ref{fig:alpha_Da} for the particle Stokes numbers $\St= 
0.1$ and $\St=1.0$. The deviation of decay rates around the mean are shown by 
the error bars.
The Damk\"ohler number is increased by increasing the number of
particles. Furthermore, the proposed modeled decay rate of the reactant molar 
fraction over the Damk\"ohler 
number according to Eq. \ref{eq:alpha_Da} is shown as the solid curve, with the 
two limiting decay rates $\alpha_p$ and $\alpha_c$ for small and large 
Damk\"ohler numbers respectively included as dotted and dashed lines.
In Fig. \ref{fig:alpha_Da} it can be seen that for cases with low 
Damk\"ohler numbers, the decay rate as predicted by DNS is 
indeed
proportional to the Damk\"ohler number and follows clearly $\alpha_p$, but 
significant deviations from the 
linear increase are observed quite early.

The decay rate begins to deviate from the linear increase (as given by 
$\alpha_p$)
for Damk\"ohler 
numbers as small as 0.1. for the given cases. For 
higher Damk\"ohler numbers the decay rate approaches the flow field dependent 
decay rate $\alpha_c$ asymptotically. The modeled decay rate $\alpha (\Da)$ 
as defined by Eq. \ref{eq:alpha_Da} fits 
the decay given by DNS rather well. 
Moreover, it is observed that the value of $\alpha_c$ is 
lower for a higher particle Stokes number.
The variance in the decay rates is higher for higher Stokes numbers, and 
this effect increases in strength for higher Damk\"ohler numbers.

\begin{figure}[H]
\includegraphics[width=.4\textwidth]{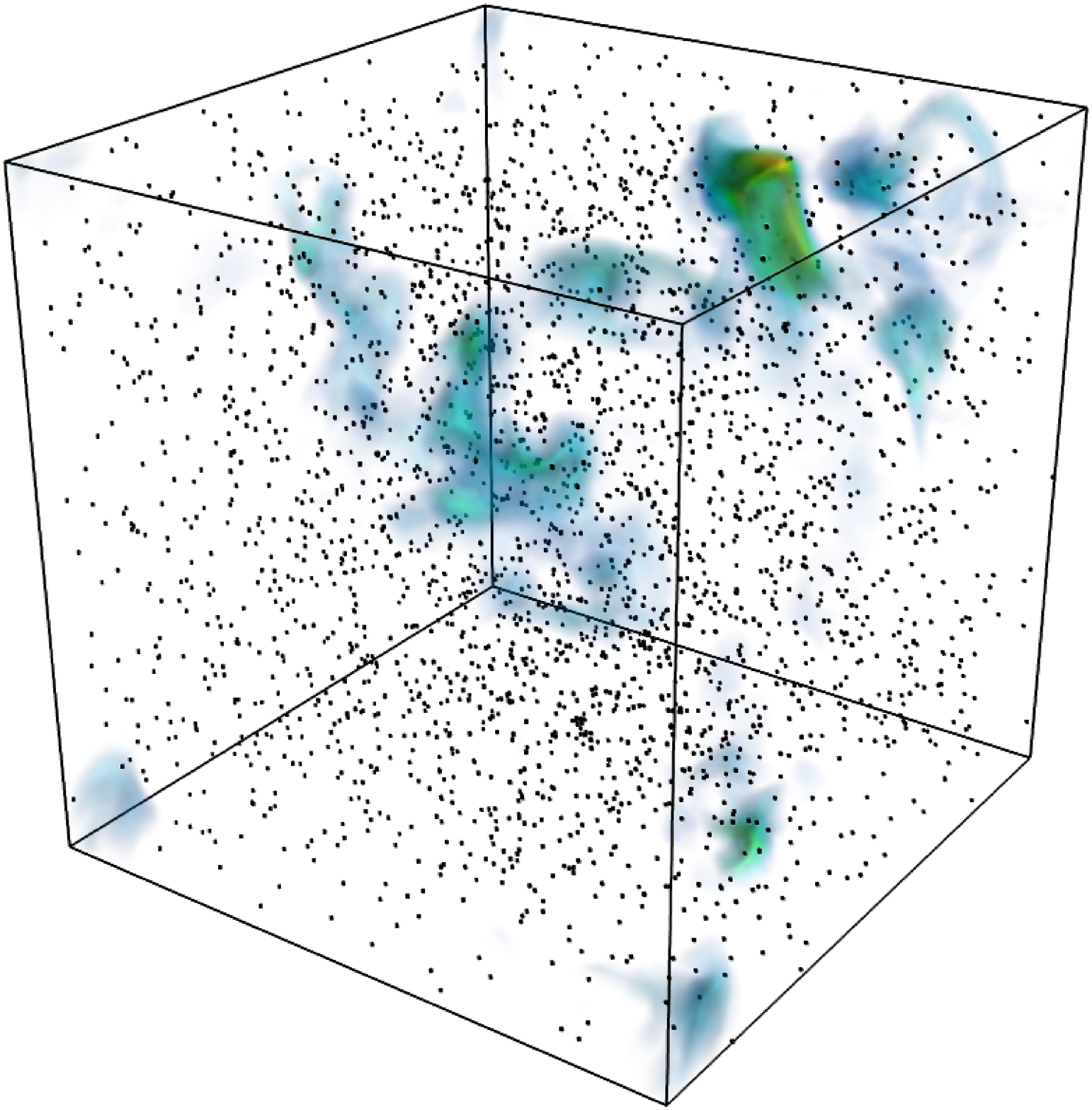}
\includegraphics[width=.4\textwidth]{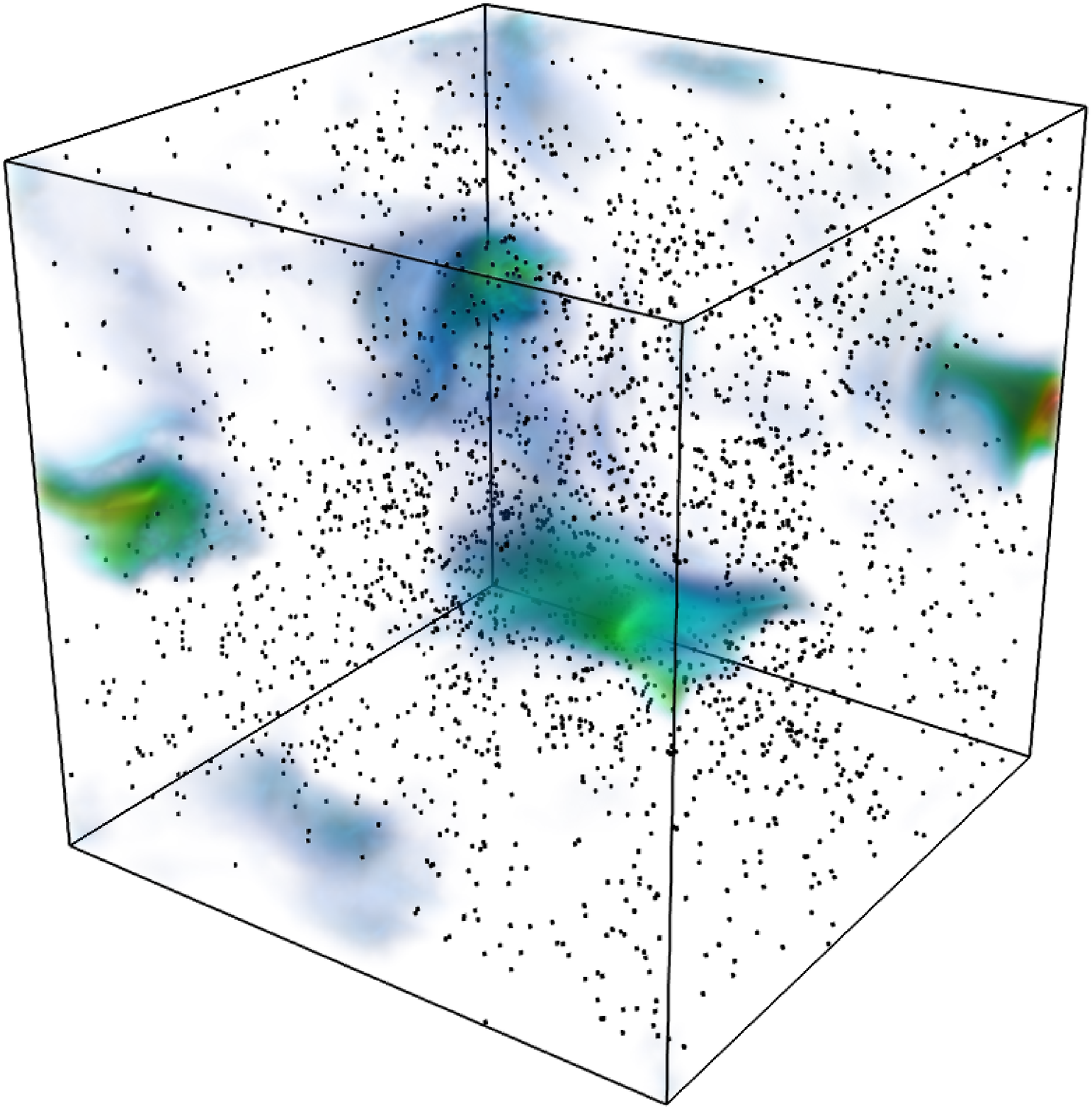}
\caption{3D plot of the domain with every 300th particle and the 90th 
percentile of the reactant concentration.}
\label{fig:box}
\end{figure}

Figure \ref{fig:box} shows a snapshot of the position of every 300th particle 
(dots) 
and the upper 90\% percentile of the reactant concentration (shaded 
areas) 
for particle 
Stokes numbers  $\St= 0.1$ and $\St=1.0$. For Stokes number of 0.1 the 
particles 
tendency to cluster is not visible and the pockets of high reactant
concentration are small. Larger areas devoid of particles can 
be seen at the higher Stokes number, as well as larger volumes of 
remaining reactant. 
Overall, the particles in the $\St=1.0$ runs show stronger 
large scale clustering than for the $\St=0.1$ runs.

\begin{figure}[H]
\begin{minipage}{30mm}
\centering\includegraphics[width=.87\textwidth]{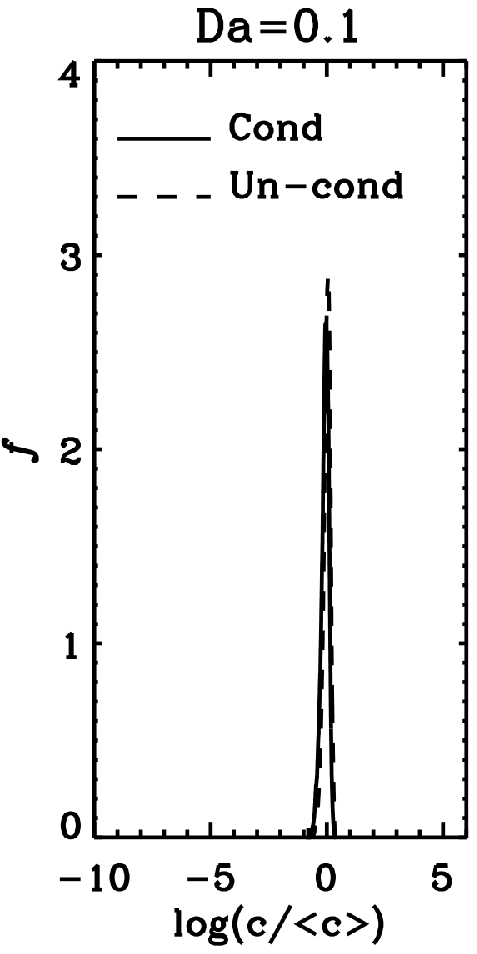}
\end{minipage}
\begin{minipage}{30mm}
\centering\includegraphics[width=.99\textwidth]{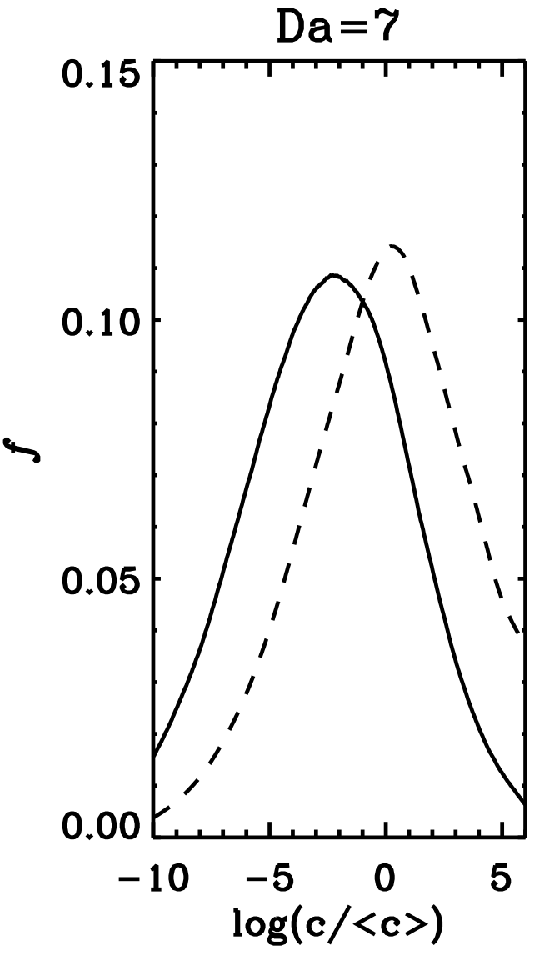}
\end{minipage}
\caption{Probability density function of logarithm of the reactant 
concentration 
$c$}
\label{fig:ml}
\end{figure}

The probability density, $f$, for a given reactant concentration $c$ is shown 
for two different Damk\"ohler numbers in Fig. \ref{fig:ml}. The dashed 
lines represent the probability averaged over the entire domain,
while the solid line represent the probability at the position of the 
particles. 
For a small Damk\"ohler number of 0.1, it is seen that the distribution is very 
narrow, and that the 
probability constraint on the particle position is nearly identical 
to the probability of all the fluid elements. This means that the reactant 
concentration is fairly homogeneous, and that it is not affected by the 
instantaneous position of the particles, i.e. that particle clustering does not 
influence the reactant distribution in a significant way. For higher 
Damk\"ohler 
numbers the distribution is broadening, which means that the reactant 
distribution is becoming less homogeneous as the importance of the particle 
clustering is increasing. One can conclude that for large Damk\"ohler  numbers 
the reactions inside the particle clusters are so fast compared to the lifetime 
of the cluster that the interior of the clusters is essentially always drained 
of reactants. This means that the reactions are happening 
at the external surface of the clusters, which resembles how reactants are 
converted at the external surface of solid objects. This in turn supports the 
assumption underlying the derivation of the asymptotic limit of $\alpha_c$.

\begin{figure}[H]
\centering\includegraphics[width=.45\textwidth]{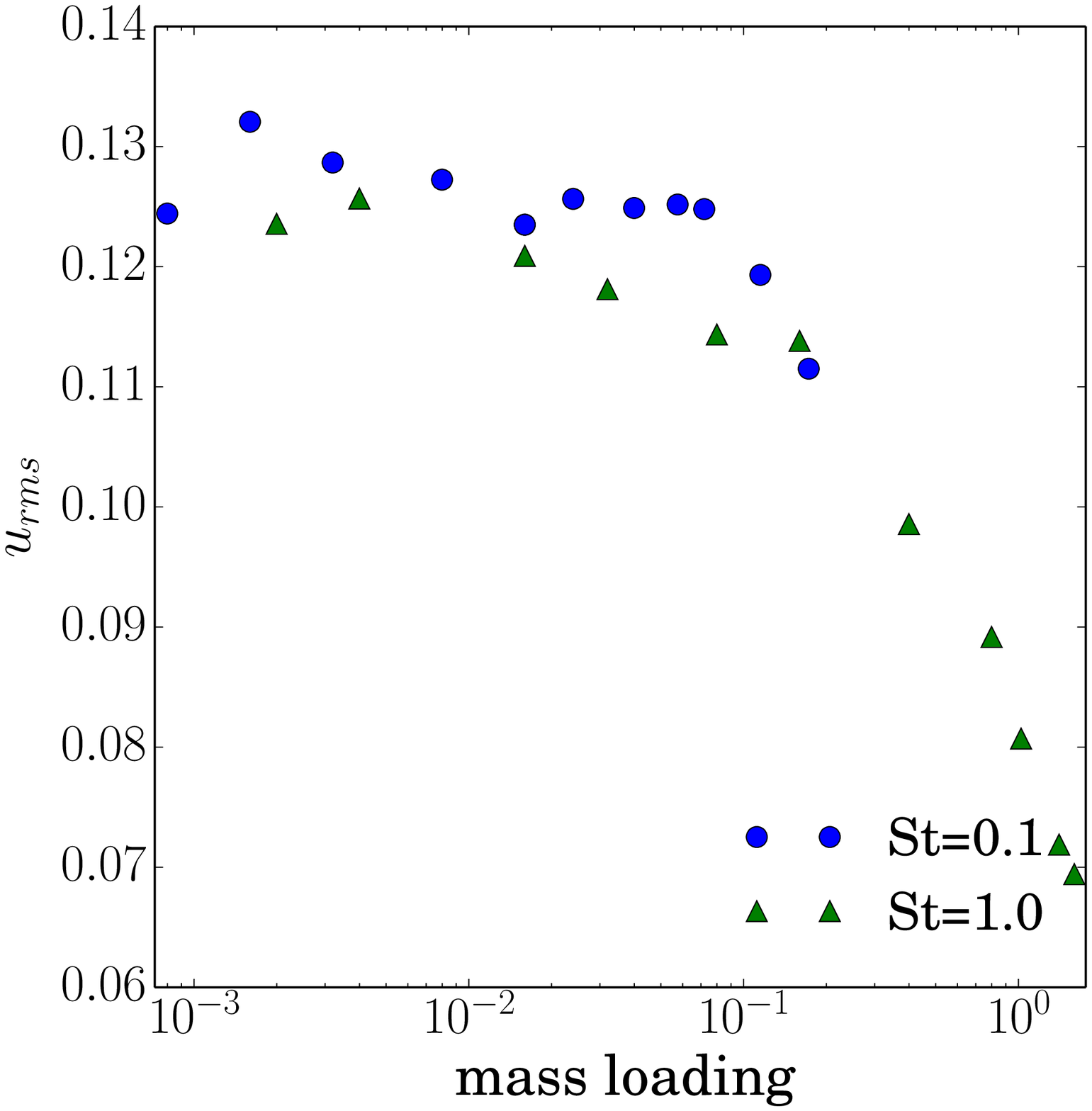}
\centering\includegraphics[width=.45\textwidth]{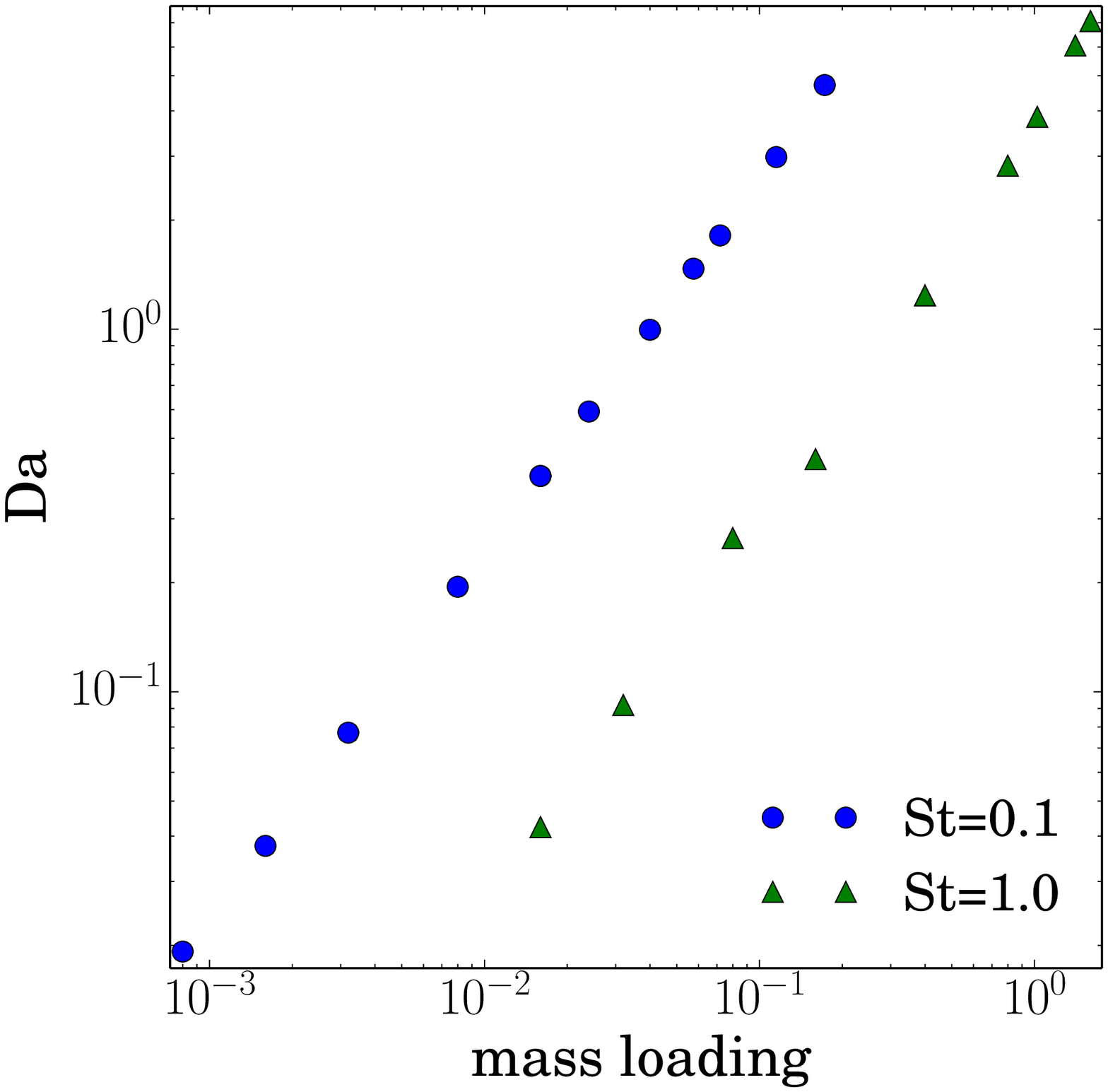}
\caption{Root mean square of velocity and Damk\"ohler number as function of 
mass loading $M_l=M_p/M_f$ for two different Stokes numbers.}
\label{fig:mlda}
\end{figure}

The mass loading $M_l$ in the simulations is defined as the ratio of
the total mass of the particles $M_p=\sum_im_{p,i}$ to the total mass
of the fluid $M_f=V\mean{\rho}$ such that $M_l=M_p/M_f$. The influence
of the mass loading on the turbulent velocities is shown in Fig. \ref{fig:mlda}.

In the upper panel the turbulent 
velocity $u_{rms}$ is shown as a function of mass loading. The mass loading 
does 
not
seem to have any significant effect on the turbulent velocity for low mass 
loadings, but for larger values of mass loading the turbulent velocity 
is significantly affected. This change in turbulent intensity will 
affect the cluster shape and in turn the surface area $\bar{A}_c$. 
For very high Damk\"ohler numbers, this may reduce the reactivity of the 
cluster 
even more than estimated by $\alpha_c$.
In the lower panel of Fig. \ref{fig:mlda}, 
the Damk\"ohler number is plotted as a function of 
the mass loading for the two different Stokes numbers. It can be observed
that the larger Stokes numbers require higher mass loadings in order to 
obtain the same Damk\"ohler number. 

\section{Conclusions}

The effect of particle clustering due to flow turbulence on the reaction rate 
of 
heterogeneous solid-fluid reactions is studied in a simplified setup. The 
particles are assumed to act like catalysts, and a simple one step reaction with 
the gas phase 
reactant on the particle surface is assumed to be fast and isothermal.
With this simplified setup it is possible to 
analytically show that for small Damk\"ohler numbers the overall 
reactivity is found not to be affected by turbulent clustering. However, for 
large Damk\"ohler numbers, the reaction rates are fast compared to the lifetime 
of the particle clusters. Hence, the effect of the clusters on the overall 
reaction rate in the domain becomes important decreasing the overall reaction 
rate. This effect is stronger for higher Stokes numbers. 

A simplified model that gives the reactant decay rate as
a function of the turbulent and chemical time scales (see Eq. \ref{eq:alpha_Da} 
) is proposed. The predictive
quantitative abilities of the presented model depend on a good representation 
of the shape, size and number density of the particle clusters.
These aspects of the cluster formation is generally
not yet properly understood
and further work to understand 
the shape and number density of particle clusters depending on flow field 
variables is needed \citep{Calzavarini2007}. Furthermore, equivalent 
investigations using more realistic heterogeneous reaction schemes and fluid 
phase physics will have to follow.

\section*{Acknowledgements}
The research leading to these results has received funding from the
Polish-Norwegian Research Programme operated by the National Centre
for Research and Development under the Norwegian Financial Mechanism
2009-2014 in the frame of Project Contract No Pol-Nor/232738/101/2014.
This work was supported by the grant "Bottlenecks for particle growth in turbulent 
aerosols” from the Knut and Alice Wallenberg Foundation, Dnr. KAW 2014.0048 and
by grant from Swedish Research Council (Dnr. 638-2013-9243).
NELH and DM also acknowledge the Research Council of
Norway under the FRINATEK grant 231444.

\bibliography{arxiv_red.bib}

\end{multicols}
\end{document}